\documentclass{article}
\usepackage{epsfig}
\usepackage{graphicx}
\usepackage{amsmath}
\usepackage{color}
\usepackage{enumerate}

\def\beq{\begin{equation}}
\def\eeq{\end{equation}}
\def\bea{\begin{eqnarray}}
\def\eea{\end{eqnarray}}
\def\bean{\begin{eqnarray*}}
\def\eean{\end{eqnarray*}}

\begin{document}

\title{Black holes and Boyle's law --- the thermodynamics of the cosmological constant}

\author{{Brian P. Dolan}
\footnote{{\em e-mail:} {\tt B.P.Dolan@hw.ac.uk}}
\\
[5mm]{\em Department of Mathematics, Heriot-Watt University}\\ 
{\em Colin Maclaurin Building, Riccarton, Edinburgh, EH14 4AS, U.K.}\\
[5mm]
{\em Maxwell Institute for Mathematical Sciences, Edinburgh, U.K.}\\
}

\maketitle
\begin{abstract}

When the cosmological constant, $\Lambda$,
is interpreted as a thermodynamic variable
in the study  of black hole thermodynamics a very rich structure emerges.
It is natural to interpret $\Lambda$ as a pressure and define the
thermodynamically conjugate variable to be the thermodynamic volume
of the black hole (which need not bear any relation to the geometric
volume). Recent progress in this new direction for black hole
thermodynamics is reviewed.

\medskip


\medskip

\rightline{PACS nos: 04.60.-m;  04.70.Dy }
\end{abstract}

\section{Introduction}

If asymptotically flat space-time contains a black hole
then a certain volume of space is hidden from an observer at infinity by the
event horizon.  The volume of space available is less because
the black hole excludes a volume that is inaccessible to the observer. 
It is however not immediately obvious how to define the
volume of a black hole in any geometric sense: for example in the line element
for a Schwarzschild black hole, described by Schwarzschild 
co-ordinates $(t,r,\theta,\phi)$, the \lq\lq radial'' co-ordinate 
$r$ is time-like inside the event horizon at $r=r_h$, 
so integrating the areas of constant $t$ shells of radius $r$ from 
$r=0$ to $r=r_h$ seems nonsensical from the point of view of the space-time geometry of
a black hole, because $r$ represents a time for $r<r_h$.
For black holes in the presence of a cosmological constant $\Lambda$
there is the added ingredient
of the vacuum energy: if a volume of space is hidden from an observe
at infinity, is the corresponding vacuum energy also hidden from them?

To avoid problems with a cosmological horizon, consider for the moment
the case of asymptotically anti-de Sitter space-time (AdS).  Then there
is a negative vacuum energy density $\epsilon=\Lambda<0$.
If a volume could be defined and a 
black hole with a volume $V$ were present then a negative energy
$\epsilon V$ would be excluded from the point of view of an observer at infinity.
Does this mean that positive energy is available to them?

Suppose we could define a volume for a black hole in some meaningful way,
then the total energy $U$ of the black hole would be not just its mass
(as defined by an asymptotic observer at infinity, 
the ADM mass \cite{Arnowitt:1960adm}-\cite{Arnowitt:1960adm3} or its equivalent in asymptotically AdS space-time \cite{Henneaux:1985}) but the mass plus the vacuum energy
hidden behind the event horizon,
\beq U=M+\epsilon V.\eeq
Since the equation of state for a cosmological constant, 
$\Lambda=-8\pi G_N P$, is that
the pressure $P=-\epsilon$, we have 
\beq M=U+PV.
\label{eq:enthalpy}
\eeq

In the thermodynamics of black holes the mass is usually interpreted
as the thermal energy and the cosmological constant is taken to be fixed.
The first law, in its simplest form, is then
\beq dM=T dS,
\label{eq:dMTdS}
 \eeq
where $T$ is the Hawking temperature and $S$ the Bekenstein-Hawking entropy.
But if the internal energy is $U$, then $M$ in equation (\ref{eq:enthalpy}) is the enthalpy
$H(S,P)$ of the thermodynamic system, the Legendre transform of $U(S,V)$,
\beq M=H(S,P)=U(S,V)+PV.\eeq
This interpretation of the black mass as enthalpy in a thermodynamic sense
was first suggested in \cite{Kastor:2009}, based on scaling arguments and
the Smarr relation \cite{Smarr:1973}.

The idea that the cosmological constant might be considered as a variable
has been suggested many times over the years, the earliest example seems to be \cite{Henneaux:1985}, \cite{Henneaux:1984}-\cite{Teitelboim:1985} and it has re-surfaced many times
\cite{Sekiwa:2006}-\cite{Ma:2013eaa}.

When the cosmological constant is fixed, we still have (\ref{eq:dMTdS}),
but if the cosmological constant  is varied infinitesimally this becomes
\beq d M = T dS + V d P, 
\label{eq:dMTdSVdP}
\eeq
where, by definition,
\beq
V=\left.\frac{\partial H}{\partial P}\right|_S.
\label{eq:ThermoV}
\eeq
This gives a thermodynamic definition of the black hole volume \cite{Kastor:2009}.
Note that this has no a priori relation to any notion of geometric
volume, it is a purely thermodynamic concept, 
we shall therefore define $V$ in (\ref{eq:dMTdSVdP}) 
to be the {\it thermodynamic volume} \cite{Dolan:2011gh}.
As already mentioned the definition of a geometric volume of a black hole is 
not straightforward, but there is
a number of suggestions in the literature, \cite{Hayward:1997jp}-\cite{Ballik:2013uia} one of which \cite{Parikh:2005qs} does actually
co-incide with the thermodynamic volume in asymptotically flat-space times, but not otherwise.

The idea of varying the cosmological constant is sometimes criticised on the basis that $\Lambda$ dictates the asymptotic form of the metric and it does not make sense to compare space-times with different asymptotic behaviour. But opening our minds and embracing the thermodynamic potential as a real physical entity in itself, putting aside questions about the underlying space-time 
which produced it,
gives a new perspective on the physics.  That is the inference of the chain of ideas presented here: the thermodynamic potential is obtained by integrating out the underlying space-time and, having done so, we are asked to forget where the potential came from and just ask what physics it gives us.  With this attitude in mind we should be careful about how phrases such as \lq\lq varying the cosmological constant'' are interpreted, we do not mean that $\Lambda$ becomes a function of time but rather we wish to ask the question, \lq\lq what would the physics look like if $\Lambda$ were different?''

In this review recent progress in the understanding of thermodynamic volumes will be described.  It is traditional to highlight the similarities between black holes and ordinary thermodynamic systems, but there are also some very important
differences and \S\ref{sec:thermodynamics} gives a quick summary of the
four laws of black hole thermodynamics with a discussion of some of the aspects
that differ from more usual systems. \S\ref{sec:4D} describes work to date
on including $\Lambda$ as a thermodynamic variable in $4$-dimensional black holes, \S\ref{sec:BTZ} covers the 3-dimensional case of asymptotically AdS BTZ 
black holes and \S\ref{sec:higherD} describes the rich structure of higher dimensional black holes, including triple points and re-entrant phase transitions.
\S\ref{sec:AdSCFT} is devoted to the cosmological constant in the boundary field
theory of the AdS/CFT correspondence and gauge/gravity duality.

\section{The four laws of black hole thermodynamics \label{sec:thermodynamics}}

The laws of black hole thermodynamics are usually stated as follows
(see {\it e.g.} \cite{Wald:1984}).

\medskip
\begin{enumerate}[(0)]

\item[(0)] The surface gravity $\kappa$ is constant on the event horizon.

\medskip

With the Hawking temperature \cite{Hawking:1974}-\cite{Hawking:1976}, $T=\frac{\hbar \kappa}{2 \pi}$, this
associates a unique temperature to the black hole in a way that agrees
with our intuitive feeling for thermodynamic systems, that the temperature is
the same everywhere in the system.

\medskip

\item[(1)] For a black hole of mass $M$, angular momentum $J$ and electric charge $Q$,
\beq dM = T\,dS + \Omega\, dJ + \Phi\, dQ
\label{eq:dM}\eeq
where the entropy $S=\frac{1}{4} \frac{A}{\hbar G_N}$ 
is proportional to the area of the event horizon \cite{Hawking:1974}-\cite{Bekenstein:1973}.

\medskip

This identifies the internal energy, 
$U(S,J,Q)$, with the mass and $T=\frac{\partial U}{\partial S}$ but, as
argued in the introduction, this is only consistent for asymptotically flat
black holes.  When there is a non-zero cosmological constant this should be modified to 
\beq dM = T\,dS  + V\,dP  + \Omega\, dJ + \Phi\, dQ,\eeq
where $M$ is the enthalpy.  Of course the first law is usually written in terms
of the thermal energy 
\beq \label{eq:UPV}
U=M-P V\eeq 
and (\ref{eq:dM}) is equivalent to \cite{Kastor:2009,Dolan:2011gh,Caldarelli:2000}

\beq \label{eq:1stLaw}
d U = T d S - P d V + \Omega d J + \Phi d Q\eeq
which reduces to the usual expression for black holes \cite{Wald:1984} when $P=0$.
An earlier version of (\ref{eq:1stLaw}) was proposed in \cite{Padmanabhan:2002}
for spherically symmetric space-times, but with $U$ replaced with $M$.
A $PdV$ term in the first law was also considered in the context of 
cosmological horizons and inflation, with no black hole,
in \cite{Ghersi:2011ggps}, and in teleparallelism theories 
in \cite{Maluf:2012na}-\cite{Castello-Branco:2013iza}.
\medskip

\item[(2)] The area of the event horizon never decreases (in any classical process).

\medskip

This relates to the identification of the entropy as being proportional to the
area.  Of course it is well known that Hawking radiation can result in the area
of the event horizon decreasing, but when the entropy of the emitted radiation
is included in the balance the total entropy of the Universe does not decrease.

\medskip

\item[(3)] It is not possible to achieve $\kappa=0$ in a finite number of steps.

\end{enumerate}

\noindent While the above identifications are very nice and resonate intuitively with
thermodynamic principles it is also important to appreciate the differences
between the thermodynamics of black holes and more standard systems.
Some of these are:

\begin{enumerate}

\item In an ordinary system the temperature is not only a property of the system
as a whole, it is also a property of parts of the system.  A macroscopic
volume of gas at a given temperature can be divided into two parts, each half
the original volume, and the parts will have the same temperature as the
original (this is the intensive nature of temperature).
This is not true for black holes, a black hole cannot in general
be broken up into two pieces each with the same temperature as the original.

\item In a thermodynamic system in $d$ space dimensions all extensive
variables scale the same way.  If lengths change by a factor $\lambda$
then the volume changes as $V\rightarrow \lambda^d V$, the entropy as $S\rightarrow \lambda^d S$, the particle number (or charge) as $N\rightarrow \lambda^d N$,
and the internal energy $U(S,V,N)$ as $U\rightarrow \lambda^d U$.
Scaling arguments then imply \cite{Callen:2006} that the Gibbs free energy
\beq
G(T,P,N)  = U -T S + PV = N \mu
\label{eq:Euler}
\eeq  
is equal to $N\mu$ where $\mu$ is the chemical potential.
Since, by definition,
\beq
d G=-S d T + V d P + \mu d N
\eeq
equation (\ref{eq:Euler})
gives the Gibbs-Duhem relation
\beq
\mu d N= - S d T + V d P.  
\label{eq:Gibbs-Duhem}
\eeq
A complete Legendre transform to intensive variables trivially gives zero,
\beq
\Xi(T,P,\mu): = G - N\mu=0.
\label{eq:XiVanishes}
\eeq
There are even axiomatic approaches to thermodynamic theory that elevate
homogeneous scaling to the status of an axiom \cite{Lieb:1999ly}.

For black holes the scaling of the thermodynamical variables (using Newton's constant to convert length into mass and $c=1$) is

\medskip

\setlength{\tabcolsep}{15pt}
\centerline{\begin{tabular}{| l| c| c| }
\hline
Thermodynamic Variable & Dimension \\
\hline
 Mass, $M$   &  $D-3$ \\
  Entropy, $S$ (area)  & $D-2$ \\
  Angular momenta, $J^i$ & $D-2$ \\
 Volume, $V$ & $D-1$ \\
 Electric Charge, $Q$ & $D-3$\\ 
 Temperature,  $T$ & $-1$ \\
 Angular velocity,  $\Omega_i$ & $-1$ \\
 Pressure, $P$ ($\Lambda$) & $-2$ \\
 Electric potential, $\Phi$ &  $0$ \\
\hline
\end{tabular}}

\bigskip
Scaling then gives the Smarr relation \cite{Smarr:1973},
\beq
(D-3)M= (D-2)TS + (D-2)\mathbf{\Omega}.
\mathbf{J} -2 P V + (D-3) \Phi Q
\label{eq:Smarr}
\eeq
(the notation $\mathbf{\Omega}.\mathbf{J}$ is used because there can be
more than one angular momentum in five or more dimensions).
In particular the Gibbs-Duhem relation in the infinitesimal 
form (\ref{eq:Gibbs-Duhem})
has no analogue in black hole thermodynamics. It would be an interesting 
project to develop an axiomatic approach to black hole thermodynamics
based on inhomogeneous scaling.

\item In black hole thermodynamics it is in fact crucial 
that the complete Legendre
transform to intensive variables, $\Xi(T,P,\Omega,\Phi)$ is not zero.  
This is because $\Xi$ is related
to the Euclidean action $I_E$ by \cite{Gibbons:2004asi}
\beq
I_E = \beta \Xi.
\eeq

In ordinary thermodynamics, for a system of $N$ non-interacting independent
particles, the partition function factorises as $Z=z^N$, where $z$ is the
partition function for a single particle. The analogue of this in black hole
thermodynamics would presumably be some kind of multi-centre black hole soliton, 
a combination of $N$ single black hole solutions, such that the Euclidean
action of the multi-centre solution is $N$ times the Euclidean action of
a single black-hole, {\it i.e.} the Euclidean actions would need to be additive. But the non-linearities of General Relativity make this difficult to achieve.
If a solitonic multi-centre black hole with non-vanishing additive action were to exist it would be the gravitational analogue of a free particle in ordinary thermodynamics.

\item The Hawking temperature of a Schwarzschild black hole is inversely
proportional to the mass, $T=\frac{\hbar}{8\pi G_N M}$, resulting in the
famous negative heat capacity of black holes, a black hole in empty space
will radiate and lose mass thus increasing its temperature making it radiate
even more.  This signals an instability 
that would normally make any such states unobservable.  
In ordinary thermodynamics a negative heat capacity would usually be considered to be unachievable in a real macroscopic system --- in  principle a fluctuation could take 
a system into such a state, but it would be so fleeting as to be unobservable.
But for a solar mass 
black hole the Hawking temperature is of the order $10^{-8}K$ and the lifetime
many orders of magnitude greater than the age of the Universe, hence such black holes are expected to persist as real objects for a long time. 

It makes sense to consider negative
heat capacity objects in black hole thermodynamics.
While it is possible to stabilise them, {\it e.g.} by introducing a negative
cosmological constant \cite{Hawking:1983}, it is not necessary to do so in order to discuss their thermodynamics.

\item The instability of asymptotically flat black holes is a persistent feature in any dimension. It was shown in \cite{Dias:2010eu} that all asymptotically flat rotating, electrically neutral, 
black holes are unstable and this was extended to include the charged 
case in \cite{Dolan:2014lea}. The proof is a nice example of one of the
differences between black hole thermodynamic and ordinary thermodynamic systems.
Local thermodynamic stability is determined by the convexity properties of thermodynamic potentials, see {\it e.g.} Ref. \cite{Callen:2006}.
In the entropy representation complete stability requires that
the entropy be a concave function of its arguments
and, conversely, in the energy representation the internal energy should be
a convex function of its arguments, which are all extensive.  Thus,
for a black hole including rotation and charge, we have $U(S,V,J,Q)$. 
Denoting the extensive variables
by $X^A$ complete local thermodynamic stability requires that the Hessian
\beq
U_{AB}=\frac{\partial U}{\partial X^A \partial X^B}
\eeq 
be a positive definite matrix, in the sense that all its eigenvalues are
positive.  

But it was shown in \cite{Dolan:2014lea}, using the Smarr relation
(\ref{eq:Smarr}), that there exists a negative norm vector 
of the bi-linear form $U_{AB}$ when $PV\le 0$, 
regardless of the values of $J$ and $Q$, and this vector was
explicitly constructed as a linear combination of the $X^A$.
For fixed $J$ and $Q$ the norm of the vector becomes positive only if\,\footnote{It is assumed that $\mathbf{\Omega}.\mathbf{J}\ge 0$.  If it is not then the moment of inertia must have at least one 
negative eigenvalue, signalling another source of instability.}
\beq PV > 2\left(\frac{D-2}{D-1}\right)(TS + \mathbf{\Omega}.\mathbf{J}),\label{eq:PVStability}\eeq
independent of the charge $Q$ (the co-efficient in (\ref{eq:PVStability})
is determined by the scaling dimensions of the extensive variables). 
In ordinary
thermodynamics the norm of the equivalent vector is always zero ---
homogeneous scaling of ordinary thermodynamics implies that there is always
a direction of neutral stability --- in black hole thermodynamics this is not the case.

\item The entropy of a solar mass black hole is essentially the area of the
event horizon measured using Planck lengths.  It is colossal, of the order of $10^{78}$. Using Boltzmann's formula (with $k_B=1$), $S=\ln({\cal W})$, 
this implies that the available number of microstates
(whatever those microstates are)\footnote{While there are models of quantum gravity that allow the entropy to be
calculated in terms of microstates \cite{Strominger:1996}, the focus of this review is on thermodynamics rather than statistical mechanics.  Not because the latter
is not important, it is even more important than thermodynamics when it is tractable, but because at the present time the microscopic theory underlying
black hole thermodynamics, presumably a theory of quantum gravity, is still not well understood.} is 
${\cal W}\approx e^{10^{78}}$.

\noindent An alternative statement of the third law, due to Nernst, is that the
entropy should vanish at zero temperature.  There is nothing pathological
about zero temperature black holes: a rotating black hole has a lower
Hawking temperature than a non-rotating one of the same mass,
\beq
T= \frac {\hbar} { 4\pi  M}\frac{\sqrt{M^4-J^2}}{(M^2 + \sqrt{M^4-J^2})},
\eeq
(with Newton's constant $G_N=1$) 
and the temperature vanishes for the maximum angular momentum $J= M^2$.
Astrophysical black holes can have angular momenta close
to the extremal value and an extremal Kerr black hole has an entropy
that is half the entropy of a Schwarzschild black hole of the same mass.
For an extremal solar mass black hole we again have an entropy of order
$10^{78}$, for a zero temperature object!
However there is no a priori contradiction with the Nernst statement of the third law
here: the Nernst statement assumes that the ground state is non-degenerate,
if the ground state is degenerate with degeneracy $g_0$, and there is a mass gap, then the entropy
would be $S=\ln g_0$ at zero temperature.  But it would seem that compatibility
with the third law implies that the ground state degeneracy of a solar mass
black hole is huge.  This already tells us something, usually a ground state degeneracy is lifted by perturbations unless a symmetry protects it --- 
Nature may be telling us that some powerful symmetry implies large
ground state degeneracies for quantum gravity that are protected against 
being lifted by perturbations.

\end{enumerate}

\section{$D=4$ space-time dimensions \label{sec:4D}}

Having laid down the rules in the previous section, this section
applies then to four dimensional space-times.

\subsection{Asymptotically anti-de Sitter black holes}

Consider first the simplest case of a neutral, non-rotating black hole.
For asymptotically AdS Schwarzschild space-time, with 
and $\Lambda=-\frac{3}{L^2}$ interpreted as a pressure $P=-\frac{\Lambda}{8\pi G_N}$,
the line element is
\beq
d s^2 = -f(r) d t^2 + \frac{1}{f(r)}d r^2 + r^2 (d\theta^2+\sin^2\theta d\phi^2),
\label{eq:AdS-Schwarzschild}
\eeq
with
\beq f(r)=\left(1-\frac{2G_N M}{r} + \frac{r^2}{L^2}\right).\eeq
The thermodynamic volume in fact evaluates to the na\"{i}ve Euclidean result
for a sphere of radius $r_h$, \cite{Kastor:2009}
\beq V=\frac {4\pi}{3}r_h^3.
\label{eq:naiveV3}\eeq 

This is no longer true for rotating black holes, the thermodynamic volume
of the asymptotically AdS Kerr metric does not look like any geometric volume.  
The line element in this case, including an electric charge, 
is \cite{Carter:1968}
\beq  \label{eq:ChargedAdSKerr}
d s^2=-\frac{\Delta}{\rho^2}\left( d t - \frac{a\sin^2\theta}{X}\, d\phi \right)^2
+\frac{\rho^2}{\Delta}dr^2 + \frac{\rho^2}{\Delta_\theta}d\theta^2 
+\frac{\Delta_\theta\sin^2\theta}{\rho^2}\left(a d t - \frac{r^2 + a^2}{X}\,d\phi \right)^2,
\eeq
where 
\bea \label{eq:MetricFunctions}
\Delta&=&\frac{(r^2+a^2)(L^2+r^2)}{L^2} -2 m r + q^2, \qquad
\Delta_\theta=1-\frac{a^2}{L^2}\cos^2\theta,\nonumber \\
\rho^2& = & r^2 + a^2\cos^2\theta, \qquad X = 1-\frac{a^2}{L^2}.
\eea
The metric parameters $m$ and $q$ are related to the mass and charge by
\beq \label{eq:MassCharge}
M=\frac{m}{X^2},\qquad Q=\frac {q} {X}\eeq
and the angular momentum is $J=a M$.

The mass (and hence the enthalpy) can be expressed in 
thermodynamic variables \cite{Caldarelli:2000} 
\beq 
M=H(S,P,J,Q)=
\frac {1}{2}\sqrt{\frac{
\left(S   + \pi Q^2 + \frac{8 P S^2}{3}\right)^2 + 
4 \pi^2\left( 1+\frac{8 P S}{3}  \right) J^2}
{\pi S}}. \label{eq:CCKmass}
\eeq
The thermodynamic volume evaluates to \cite{Cvetic:2011,Dolan:2011kl}  
\beq
V= \left.\frac {\partial H}{\partial P}\right|_{S,J,Q}=\frac{2}{3 \pi H}\left[S\left(S + \pi Q^2 + \frac{8 P S^2}{3}\right)+ 2\pi^2 J^2  \right].
\eeq
 In terms of the geometric parameters $a$, $q$ and $L$ in the
metric (\ref{eq:ChargedAdSKerr}) this is 
\beq \label{eq:Volume}
V = \frac {2\pi }{3}\left\{
\frac{(r_h^2+a^2)\bigl(2 r_h^2 L^2 + a^2 L^2 - r_h^2 a^2 \bigr) +  L^2 q^2 a^2}
{L^2 X^2 \,r_h}\right\}\ge \frac {4\pi }{3}r_h^3. 
\eeq
A geometric definition of the volume of a black hole was given in \cite{Parikh:2005qs} which in fact agrees with (\ref{eq:Volume})
for neutral black holes in the asymptotically flat limit, 
but they give different results when $L$ is finite (and so $\Lambda<0$).

The formula for the thermodynamic volume in terms of geometric variables
highlights the fact that $V$ is equal to $\frac {4\pi }{3}r_h^3$ if and only if $a=0$, {\it i.e.} for $J=0$ the volume 
reduces to the Euclidean result (\ref{eq:naiveV3}) for any $q$, they differ only for $J\ne 0$.  Comparing this with the entropy
\beq 
S=\frac{\pi(r_h^2+a^2)}{X}\ \mathop{\longrightarrow}_{a\rightarrow0}\ \pi r_h^2
\eeq
we see that the entropy and the volume are not independent when $J=0$, even
for charged black holes.  This can lead to subtleties in taking Legendre
transforms \cite{Dolan:2011kl}.

The inequality in (\ref{eq:Volume}) is interesting, since rotating black holes
necessarily deviate from spherical symmetry it implies that non-symmetric
black holes have a smaller surface to volume ratio than spherically symmetric
black holes of the same radius.  
This is contrary to our Euclidean intuition where
the geometry with the smallest surface to volume ratio is a perfect sphere,
any deviation from perfect symmetry increases the surface to volume ratio  ---
the isoperimetric inequality of Euclidean geometry.  This led the authors
of Ref. \cite{Cvetic:2011} to postulate a {\it reverse isoperimetric inequality} for thermodynamic volumes of asymptotically AdS black holes:
that the entropy inside a horizon of a given “volume” $V$ is
maximised for Schwarzschild-AdS. 

When discussing thermodynamic potentials it is important to bear in mind
what the correct control parameters are \cite{Callen:2006}.
For example the thermal energy $U(S,V,J,Q)$ is a function of volume and
\beq
P=-\left(\frac{\partial U}{\partial V}\right)_{S,J,Q}.
\eeq
To express $U$ as a function of $P$, 
\beq U(S,P,J,Q)=U\left(S,-\frac{\partial U}{\partial V},J,Q\right),\eeq
implies a differential equation whose solution requires an undetermined constant.
There is nothing wrong with writing $U(S,P,J,Q)$ as long as we remember that
some information is lost in doing so.  This is not necessary for
the case in hand though as the Legendre transform (\ref{eq:UPV})
can be performed explicitly and the internal energy determined in terms of the
natural thermodynamical variables \cite{Dolan:2011kl}
\bea \label{eq:InternalEnergy}
U(S,V,J,Q)&=&  \left(\frac {\pi}{S}\right)^3\left[ 
\left(\frac{3V}{4\pi}\right)
\left\{ \left(\frac S {2\pi}\right)\left(\frac S {\pi} + Q^2 
\right)+J^2 \right\}\vline  height 20pt  width 0pt depth 20pt
\right.\\
& & \kern 50pt \left. -|J|\left\{\left(\frac{3V}{4\pi}\right)^2-\left(\frac S \pi \right)^3\right\}^{\frac 1 2}
\left(\frac{S Q^2}{\pi}+J^2\right)^{\frac 1 2}
 \right].  \nonumber
\eea

The presence of the $PdV$ term in (\ref{eq:1stLaw}) has physical consequences,
it is possible to extract more energy in a Penrose process in asymptotically
AdS space-time than in asymptotically flat space-time.
The maximal efficiency for an electrically neutral black hole in the former 
case is \cite{Dolan:2011kl} $51.8$\%, while the value for the latter case is \cite{Wald:1984}
$1-\frac 1 {\sqrt{2}} =29.3$\%.  The corresponding figures for a charged
rotating black hole are $75$\% and $50$\% respectively. 

As mentioned above when $J=0$ the thermodynamic volume (\ref{eq:Volume})
reduces to the Euclidean result $\frac{4\pi}{3}r_h^3$, which is not independent
of the entropy $S=\pi r_H^2$.  It was observed in \cite{Johnson:2014yja}
that this has the amusing consequence that the adiabats for a Carnot cycle
are also isochoric ($V=const$) and hence are vertical lines 
in the $P-V$ plane.  In terms of the familiar adiabatic condition $PV^\gamma = const$, re-written as $P^{\frac 1 \gamma} V = const$, this is equivalent to the
statement that $\gamma=\infty$.  In this 
sense non-rotating black holes behave like
a gas of molecules with an infinite number of excitable internal degrees of 
freedom.

The introduction of the pressure as another thermodynamic variable in black hole
physics expands our view of the thermodynamic picture.  
It was shown in Ref. \cite{Caldarelli:2000} that an asymptotically AdS rotating black hole has a phase transition, when $J$ is held fixed,
between large and small black holes.  This was originally discussed in terms
of fixed $\Lambda$ and can be seen by plotting curves of fixed $J$ in the $T-S$ plane.  As $J$ is increased,
keeping $\Lambda$ fixed,
there is a critical value $J_c$ at which there is a second order phase transition, the transition is first order for $J<J_c$ and there is only one phase
for $J>J_c$, while at $J=0$ the small black holes disappear.
This phase transition, which we shall call the CKK phase transition, 
is distinct from the Hawking-Page phase transition \cite{Hawking:1983}.
The latter occurs at constant angular velocity, $\Omega$, 
when the free energy of the black hole is larger than the free energy of AdS
with no black hole, but with thermal radiation rotating with angular velocity
$\Omega$, and the black hole decays to AdS plus radiation.
Discussion of the Hawking-Page phase will be postponed to later in this section
and we shall focus on the CKK phase transition first. 

The CKK phase transition can also be pictured by fixing $J$ and varying
$P$ (or $\Lambda$).  When isotherms are plotted in the $P-V$ plane the 
similarity with the van der Waals gas becomes striking \cite{Dolan:2011kl}.
While the equation of state $T(P,V)$ cannot be obtained in closed form for non-zero $J$ 
(that would require extracting the roots of an eighth order polynomial),
a virial expansion can be performed and critical exponents extracted at the critical point.  The critical exponents $\alpha$, $\beta$, $\gamma$ and $\delta$,
are defined in the usual way (with reduced temperature $t=\frac{T-T_c}{T_c}$ and
reduced volume $v=\frac{V-V_c}{V_c}$),
\begin{itemize}
\item the specific heat diverges as $C_V \sim |t|^{-\alpha}$;
\item the jump in volume just below $T_c$, between the gaseous (large black hole)
phase, $v_>$, 
and the liquid (small black hole) phase, $v_<$, vanishes as, 
\hbox{$v_> - v_< \sim  |t|^\beta$};
\item the isothermal compressibility diverges as $\kappa_T \propto |t|^{-\gamma}$;\item at the critical temperature the pressure $|P-P_c| \propto |v|^\delta$.
\end{itemize}
For the CKK phase transition the exponents
are mean field 
\beq \alpha=0, \qquad \beta=\frac 1 2, \quad \gamma=1, \qquad \delta=3,
\eeq 
putting the asymptotically AdS Kerr black hole in the same universality class as the van der Waals gas. 
\cite{Dolan:2012jh,Gunasekaran:2012dq}
The same exponents are found in a singularity in the isothermal
moment of inertia tensor \cite{Tsai:2011gv}.

\begin{figure}[!ht]
\centerline{\includegraphics[width=8cm]{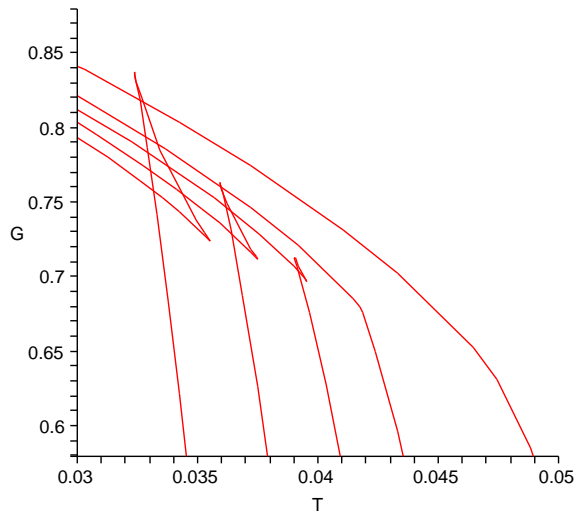}}
\caption{The Gibbs free energy in $D=4$, as a function of $T$ at various
pressures and fixed $J=1$.} 
\label{fig:Gibbs}
\end{figure}

This phase transition can be analysed in terms of the Gibbs free energy
for the system, $G(T,P,J)=M-TS$.  
This is a multi-valued function of its arguments and the
most stable phase corresponds to the lowest branch of $G$.
For an asymptotically AdS, electrically neutral,
rotating black hole $G$ is plotted in figure \ref{fig:Gibbs}, as a function of
$T$ for $J=1$ and various values of the pressures near the critical value $P_c$.
The thermodynamically stable configuration is the one of lowest $G$ and
the lower envelope of $G$ is a concave function.  For pressures below
the critical pressure there is a kink where $\left|\frac{\partial G}{\partial T}\right|_{P,J}$ is discontinuous, corresponding to a first order phase transition,
for pressures above the critical value the kink disappears.
The singularity associated with the critical point is that of a swallowtail
in the language of catastrophe theory, \cite{Sewell:1990} 
$A_4$ in the $A-D-E$ classification of singularities \cite{Arnold:1994}.

\begin{figure}[!ht]
\centerline{\includegraphics[width=8cm]{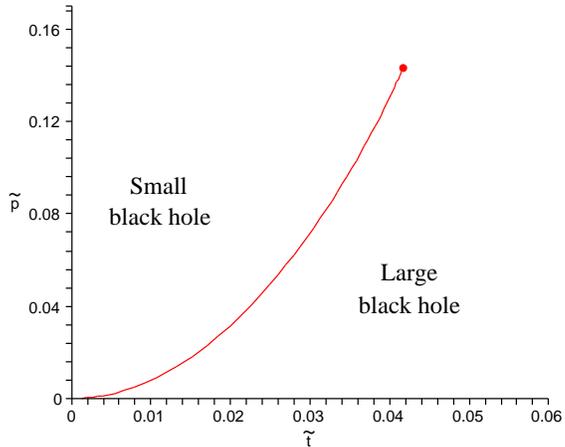}}
\caption{Co-existence curve of the CKK transition with fixed $J$,
between large and small black holes, in the $P-T$ plane.  
There is a line of first order transitions
terminating in a critical point which is a second order phase transition
with mean field exponents \cite{Gunasekaran:2012dq,Dolan:2012jh}.
The pressure and temperature are here rendered dimensionless using suitable
powers of $J$, (with $G_N=1$) $\tilde t=T\sqrt{J}$ and $\tilde p=16\pi PJ$. 
} \label{fig:P-T}
\end{figure}

When viewed in the $P-T$ plane the phase boundary of the CKK transition
re-enforces the analogy with the van der Waals liquid-gas transition.
There is a line of first-order phase transitions terminating in
a critical point where there is a second order phase transition, as shown
in figure \ref{fig:P-T}.

Electrically charged black holes show similar features,
indeed the first suggestion of van der Waals type behaviour for a black hole
was an analogy between the non-rotating charged black hole equation of 
state, $\Phi(Q,T)$ at fixed $P$, and 
the van der Waals equation, $P(V,T)$, based purely on charge and electric 
potential with no
reference to volume, \cite{Chamblin:1999ab,Chamblin:1999cd}
an analogy which was pursued in \cite{Wu:2000}-\cite{Sahay:2010tx}.
Bringing the thermodynamic volume into the picture makes this analogy even 
closer.  In the non-rotating case the equation of state can be determined exactly
\beq
P= \frac T 2 \left(\frac {4\pi}{V}\right)^{\frac 1 3} 
+\frac {1} {8\pi}\left(\frac {4\pi}{V}\right)^{\frac 2 3} 
+\frac{Q^2}{8\pi} \left(\frac {4\pi}{V}\right)^{\frac 1 3},
\eeq
and the critical point has mean field exponents, which can be seen in both in
the $Q-\Phi$ plane \cite{Niu:2011tb} and in the $P-V$ plane \cite{Kubiznak:2012wp}.

 When both $J$ and $Q$ are non-zero 
there is a line of second order phase transitions in the $J-Q$
plane encircling the origin \cite{Caldarelli:2000},
each point of which has mean field critical exponents \cite{Dolan:2012jh}.
The $J=0$ case uniquely has the property that the
thermodynamic volume equals the na{\" i}ve Euclidean volume,
and the volume is therefore not independent of the entropy,
rotation is necessary to free the thermodynamic volume from the entropy.
The critical point in the charged rotating case was studied in \cite{Belhaj:2013cva}.

Non-linear modifications of electrodynamics have also been considered in
the context of varying $\Lambda$.
For example one can replace the Maxwell action with the Born-Infeld action \cite{Born:1934,Gibbons:2003}
\beq
{\cal L}_{BI}= -b^{-4}\Bigl(\sqrt{-\det\bigl(g + b^2 F\bigr)} - \sqrt{-\det g}\Bigr),
\eeq
(where $g_{\mu\nu}$ the space-time metric, $F_{\mu\nu}$ the Maxwell field strength
and $b$ a fixed constant with dimensions of length). Charged asymptotically AdS black hole solutions of 
the  Einstein-Born-Infeld system are known \cite{Fernando:2003fk}-\cite{Cai:2004cpw} and have been analysed from the perspective adopted here
($b$ is a new dimensionful parameter which affects the Smarr relation).  
Divergences in the heat capacity on the spinodal curve\footnote{The spinodal curve is defined to be the curve, in the $T-S$ or the $P-V$ plane, on which response functions, such as heat capacity or compressibility, diverge.}  
were examined in \cite{Banerjee:2010da,Lala:2011np} and exponents 
characterising the divergence were calculated in \cite{Banerjee:2011cz}.  
At the critical point, where the two branches of the spinodal curve meet, 
the character of the divergence changes and
the critical exponents associated with solutions of the Born-Infeld action 
are again mean field \cite{Gunasekaran:2012dq}. A new feature in Born-Infeld
electromagnetism is that there is a range
of $b$ for which there can be a discontinuous jump in the value of the lowest
branch of the Gibbs free energy --- a zeroth order phase transition \cite{Gunasekaran:2012dq}.

For charged non-rotating black-holes the van der Waals type behaviour
first observed in \cite{Chamblin:1999ab,Chamblin:1999cd} was for fixed electric charge.  If the system is coupled to a charge reservoir and the electric potential fixed there is no critical point.  However if, in addition, the black hole is
given a magnetic charge a rich phase space structure re-emerges \cite{Dutta2013:djs} with large and small black hole phases and a second order phase transition.
The inclusion of a magnetic charge on the black hole is an essential ingredient
for understanding the quantum Hall effect from the AdS/CFT point of view \cite{Hartnoll2007:hk,Bayntun2011:bbdl}.

Maxwell's equal area law can be used to identify the temperature of the 
phase transition, equating free energies in the two phases determines
the temperature of the transition when it is below the critical temperature,
and this was investigated in \cite{Tsai:2011gv,Spallucci:2013aab,Spallucci:2013jja}.
This is however
one of those situations where one must be aware of the differences between the
thermodynamics of black holes and more familiar systems: in a fluid 
there is a definite ratio of the amount (molar mass) of fluid in the liquid phase to the amount in the gaseous phase at any point on the horizontal segment
of the isotherm constructed using the equal area law, 
given by the \lq\lq lever rule'', \cite{Callen:2006}
but there is only one black hole in the CKK transition and it does not
seem sensible to apportion it into a partly large black hole and partly small black hole.

Knowing the equation of state allows one to calculate a compressibility
for a rotating black hole, \cite{Dolan:2011jm} either adiabatic
\beq \label{eq:kappa_S}
\kappa_S =-\frac{1}{V} \left(\frac{\partial V}{\partial P}\right)_{S,J,Q}
\eeq 
or isothermal
\beq
\kappa_T =-\frac{1}{V} \left(\frac{\partial V}{\partial P}\right)_{T,J,Q}.
\eeq 
(Non-rotating black holes are adiabatically incompressible, 
as fixing $S$ completely fixes $V$ when $J=0$.)  This is one place where
one can connect with astrophysical black holes as the compressibility
(\ref{eq:kappa_S}) remains non-zero as $\Lambda\rightarrow 0$.
The $Q=0$ black
hole equation of state for a solar mass black hole is significantly 
stiffer than that of a neutron star of the same mass, assuming an ideal
degenerate gas of neutrons.

An adiabatic speed of sound can be defined using the usual thermodynamic 
formula, defining an average density in a natural way $\rho=\frac{M}{V}$
a speed of sound, $v_s$, can be defined through
\beq v_s^{-2}=\left.\frac{\partial \rho}{\partial P}\right|_{S,J,Q}
=1+\rho\kappa_S=1+\frac{9(2\pi J)^4}{\bigl(6 S^2 + 16 P S^3 + 3 (2\pi J)^2\bigr)^2} \ge 1,
\eeq
with equality for $J=0$, when the compressibility vanishes.
Of course $v_s$ cannot be the speed of a surface wave on the black hole,
it is probably better thought of as the velocity of a breathing mode due
to the changing volume at constant $S$.

 So far we have only discussed the properties of asymptotically AdS black
holes with fixed angular momentum, which exhibit the CKK phase transition,
and/or fixed charge, which exhibit the van der Waals type phase transition
described in \cite{Chamblin:1999ab,Chamblin:1999cd}. 
There is of course another kind of phase transition, first found by Hawking
and Page \cite{Hawking:1983}, in which the free energy of the black hole 
with a given Hawking temperature is unstable and decays to AdS space-time,
filled with radiation at the same temperature, due to competition
between the free energies.  The black hole will decay if its free energy is
greater than that of AdS with no black hole\footnote{The individual free energies
are infinite, but their difference is finite \cite{Hawking:1983},
alternatively \cite{Henningston:1998,Balasubramanian:1999} the free energy can regularized by adding suitable counter terms at large $r$ and letting $r\rightarrow \infty$, both these procedures
give the same answer.} 
and, conversely, if the 
Hawking temperature is such that the black hole as a lower free energy
than AdS with radiation at the same temperature then black holes will be
expected to spontaneously nucleate in AdS space-time. 
If the black hole has fixed non-zero charge or angular momentum then this
is not an issue because conservation of charge/angular momentum stops
any such transitions, but a rotating black hole with fixed angular velocity
$\Omega$ can decay to AdS with rotating radiation at the same temperature
(providing it is not rotating too fast --- such a rotating black hole is asymptotically conformal to a 3-dimensional rotating Einstein universe and the angular
velocity is limited if we do not want the Einstein universe to rotate faster than the speed of  light \cite{Hawking:1999}).
 
\begin{figure}[!ht]
\centerline{\includegraphics[width=8cm]{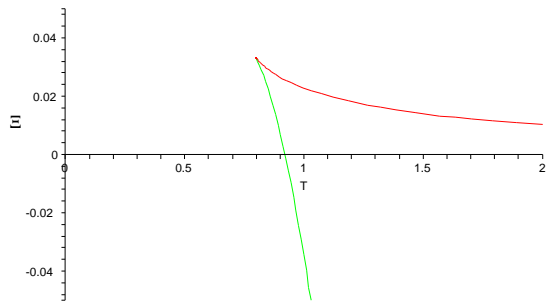}}
\caption{Free energy as a function of $T$ at fixed $P$ and $\Omega$.
The point where the lower branch crosses the $T$-axis is the Hawking-Page
phase transition.} \label{fig:Gibbs_Hawking-Page}
\end{figure}

The free energy for a black hole at constant $\Omega$,
\beq
\Xi(T,P,\Omega)= M - TS - \Omega J,
\eeq
is shown in figure \ref{fig:Gibbs_Hawking-Page} as a function of $T$, at fixed $P$. There are two branches, the upper branch is
small black holes and the lower branch large black holes, the Hawking-Page
phase transition occurs at the point where $\Xi$ becomes negative and large
black holes are more stable than AdS with thermal radiation and no black hole, for which $\Xi=0$.
There is a cusp in $\Xi$ at the point where the heat capacity diverges, but
there is no second order phase transition as one cannot
go through the cusp to lower $T$ --- the black hole solution simply does not
exist for lower $T$.

Many other standard thermodynamic quantities can be checked in various
situations: 

\begin{itemize}

\item The Clapeyron equation for first order transitions, 
relating the slope of the phase boundary in the $P-T$ plane
to the ratio of the jump in entropy to the jump in volume.
For the Hawking-Page phase transition the jump in volume is $\Delta V =V$ 
(since $V=0$ in the AdS phase) and similarly $\Delta S =S$ and the
Clapeyron equation reduces to 
\beq
\frac{d P}{d T}=\frac{\Delta S}{\Delta V} = \frac{S}{V}.
\eeq
For the Hawking-Page phase transition this evaluates to
\beq \frac{d P}{d T}=\frac{2 P S}{M},
\eeq
which is easily checked using the explicit solution.  The latent heat 
across the first order transition is \cite{Dolan:2014mnn}
\beq
L=T\Delta S + \Omega \Delta J=\frac{16 \pi^3 T^3}{(4 \pi^2 T^2 -\Omega^2)^2}.
\eeq

\item For second order transitions $\Delta V = \Delta S=0$ and the
Clapeyron equation is not directly applicable. 
The appropriate framework is then Ehrenfest's equations and these were 
investigated for charged, non-rotating black holes in \cite{Mo:2013ela}-\cite{Mo:2014wca}.  

For rotating black holes Ehrenfest's  equations are satisfied for neutral black holes at constant
$\Omega$, \cite{Banerjee:2011bms} and also for charged black holes \cite{Banerjee:2012mbr}.
 
\item The thermodynamic volume has also been explored in modified gravitational theories,
beyond the simple Einstein action, such as:\newline
quantum corrected, asymptotically safe gravity \cite{Ma:2014vxa};
$f(R)$ gravity \cite{Mo:2014lza}; and conformal gravity \cite{Xu:2014kwa}. 

\item Quasi-normal mode frequencies show a marked change near second order phase transitions between large and small black holes \cite{Tharanath:2014uaa,Liu:2014gvf}.

\end{itemize}

Another direction that has been pursued is to include the cosmological constant
as a thermodynamical variable in the application of
geometrothermodynamics to black holes \cite{Ruppeiner:2013}.
Geometrothermodynamics investigates a natural geometry associated
with the curvature of thermodynamic potentials (not to be confused with the
geometry of space-time!).
It was found in \cite{Larranaga:2014lm} that including $\Lambda$ as
a thermodynamic variable dos not affect the singularity structure of the
thermodynamic Ricci scalar. 

\subsection{de Sitter black holes}

A positive cosmological constant has subtleties associated with it that are
absent in the asymptotically AdS geometry: notably 
the presence of two horizons,
a black hole and a cosmological horizon, each with different temperatures;
the absence of a time-independent, space-like, asymptotic regime and resulting  
difficulties in defining a mass.  Nevertheless progress can be made, 
using Komar forms to define an invariant mass \cite{Dolan:2013ft} (resulting in essentially the same result as would be obtained by taking the AdS form of the mass \cite{Henneaux:1985} 
and analytically continuing to negative pressure).
One outcome of this analysis is that, if the thermodynamic volume of the Universe is fixed, the entropy of the Universe is always increased if a black hole is added.
The topic was further explored in \cite{Ma:2013gbb,Zhao:2014tsb}
and a position dependent mass 
function related to enthalpy proposed in \cite{{Bhattacharya:2013bl}}, based on
earlier work in \cite{Abbott:1982ad}.

Quintessence quintessentially involves a positive cosmological
constant and the equation of state for Reissner-Nordstr\"om black holes
in quintessence theories was shown in Ref. \cite{Li:2014ixn} 
again to have a large black hole/small black small black hole transition 
with a critical point possessing mean field exponents.

\subsection{Other $4-D$ space-times}

Following our general philosophy here one can ask questions about the
volume associated to any space-time for which the Euclidean action, $I_E$,
produces a non-trivial function depending on $P$.
The Euclidean action is related to a thermodynamic potential (Massieu function \cite{Callen:2006}) $\Xi(T,P,\Omega,\Phi)$, which is a function of purely intensive variables, via $\Xi=T I_E$ \cite{Gibbons:2005gpp}. 
For example asymptotically AdS Taub-Nut and Taub-bolt solutions were studied 
in Refs. \cite{Johnson:2014xza}-\cite{Lee:2014tma}  
while the Kerr-bolt was treated in Ref. \cite{MacDonald:2014zaa}.  A curious feature is that the 
AdS-Taub-NUT 
thermodynamic volume is negative, a result that can be understood 
in terms a \lq\lq formation process'' of Taub-NUT space-time from pure 
AdS which has the effect of increasing the volume of space-time \cite{Johnson:2014xza}.

\section{$D=3$ and the BTZ black hole \label{sec:BTZ}}

Black holes in 3-dimensions have been a source of inspiration in gravitational research ever since the discovery of the BTZ solution \cite{Banados:1992btz,Banados:1993btz}.  They provide theoretical insights, both classical and quantum,
into the physics of black holes which are much harder to analyse in 4-dimensions. For a review of BTZ black holes see \cite{Carlip:1995}.

The thermodynamic volume of non-rotating asymptotically AdS BTZ black holes was calculated in \cite{Dolan:2011gh} and, as for other dimensions,
it gives the Euclidean geometric result: in this case the area of a disc of radius $r_h$, $\pi r_h^2$. In three dimensions however one can make progress in
including quantum gravity corrections to the free energy of the black hole \cite{Maloney:2010mw}. The thermodynamic volume arising from the quantum corrected
free energy was calculated in \cite{Dolan:2011gh} and it was found that the
pressure was reduced at low volumes compared to the classical equation of
state. This was the first example of a thermodynamic volume that differed
from the na{\"i}ve Euclidean geometry result.  For the rotating BTZ there
is no CKK critical point in 3-dimensions \cite{Gunasekaran:2012dq}, a conclusion
which extends to non-linear Born-Infeld electrodynamics \cite{Zou:2013owa}.
When scalar hair is included however a van der Waals type phase transition was
reported in \cite{Belhaj:2013ioa}.
Also when the Einstein action is modified to include a Chern-Simons term
with torsion, as is very natural for the asymptotically AdS BTZ black hole,
there are divergences in the heat capacity and the form of these
divergences on the spinodal curve were investigated in \cite{Ma:2014tka}.

\section{$D>4$ space-time dimensions \label{sec:higherD}}

In more than 4-dimensions a black hole can have more than one angular momentum, due to the fact that
the rank of the rotation group in $(D-1)$ space dimensions is greater than one
and the number of independent angular momenta is equal to the rank $r$ of $SO(D-1)$, {\it i.e.} $\frac{D-2}{2}$ in even dimensions and  $\frac{D-1}{2}$ in odd dimensions.

For charged static black holes in more than 4-dimensions the situation is similar to $4-D$, there is a critical point with mean field exponents.
\cite{Gunasekaran:2012dq, Belhaj:2012bg,Mo:2014mba}
The observation that the thermodynamic volume agrees with the na\"{i}ve volume
of a perfect sphere in Euclidean space also generalises to $D$ space-time dimensions \cite{Dolan:2011gh},
the thermodynamic volume evaluates to 
that of a $(D-2)$ dimensional sphere of radius $r_h$ in $(D-1)$ dimensional
Euclidean space.
As in $D=4$ this is no longer the case for rotating black holes \cite{Cvetic:2011} and the reverse isoperimetric inequality conjecture extends to all $D\ge 4$.

In $D>4$ the phase space structure of black holes rotating with constant angular
momentum is much richer than in 4-dimensions. 
In addition there are black holes with event horizons which do not
have the topology of a $D-2$ dimensional sphere, such as 
black rings \cite{Emparan:2001wn} and black saturns \cite{Elvang:2007rd},
but the thermodynamic volume of such objects has yet to be investigated.

Even when all but one of the angular momenta are set to zero 
(singly spinning black holes) the structure of the thermodynamic phase space can be
more involved than $D=4$.
In all dimensions $D\ge 6$ singly spinning black holes can exhibit the phenomenon of
re-entrant phase transitions where phases $A$ and $B$ alternate as $A-B-A$
when a single parameter is varied \cite{Altamirano:2013akm}, in this case
varying the temperature, at fixed pressure and angular momentum,
with phase $A$ being large black holes and $B$ small black holes.

When more than one angular momentum is non-zero there are more possibilities. 
In $D=6$ the state of rotation is described by two angular momenta, $J_1$ and $J_2$, and the structure of the phase space depends on the ratio ${J_1}/{J_2}$.
In addition to the CKK transition, familiar in 4-dimensions,
there can now be multiple first-order transitions between three
different phases: small, intermediate size and large black hole phases,
as well as triple points \cite{Altamirano:2014tva,Altamirano:2013uqa}.
 
The adiabatic compressibility of higher dimensional asymptotically AdS rotating black holes, specifically asymptotically AdS Myers-Perry black holes, \cite{Myers:1986mp}-\cite{Gibbons:2004glpp2},
was evaluated in \cite{Dolan:2013dga}.  It is positive and bounded above for $P>0$, but can diverge in the asymptotically flat limit $P\rightarrow 0$.
The speed of sound associated with a Myers-Perry 
(asymptotically flat) black hole has an elegant expression in terms of quadratic and quartic Casimirs of $SO(D-1)$ when $\Lambda \rightarrow 0$: 
defining dimensionless angular momenta as the angular momenta per unit entropy, ${\cal J}_i=\frac{2\pi J_i}{S}$ with $i=1,\ldots,r$, the speed of sound, as $\Lambda \rightarrow 0^-$, evaluates to 
\beq
v_s^2=\frac{1}{(D-2)} 
\frac{\bigl(D-2 + \sum_i {\cal J}_i^2\bigr)^2}{\bigl(D-2 + 2\sum_i {\cal J}_i^2 + \sum_i {\cal J}_i^4\bigr)},\label{eq:SpeedOfSound} 
\eeq
which lies in the range $\frac {1}{D-2} < v_s^2 \le 1$. Non-rotating black holes
are incompressible, with $v_s^2=1$, while the lowest speed is given by
$v_s^2\rightarrow \frac{1}{D-2}$, when one\footnote{Which can happen in $D\ge 5$, a phenomenon termed \lq\lq ultra-spinning'' black holes.
This is associated with a divergence in the moment of momentum tensor, the angular velocity remains finite.} 
or more ${\cal J}_i\rightarrow\infty$.  This is still higher than the speed of sound in a photon gas in $D$-dimensions, which is $v^2=\frac{1}{D-1}$.

Local thermodynamic stability was discussed in \S\ref{sec:thermodynamics}.
The stability of a thermodynamic system depends on what external constraints 
are applied \cite{Dolan:2014lea}.
Including the cosmological constant as a thermodynamic variable raises questions
about local thermodynamic stability over and above the usual considerations
of positive heat capacity and positivity of the moment of inertia.
The thermal energy $U(S,V,J_i,Q)$ is now a function of $r+3$ variables, where $r$ is the rank of $SO(D-1)$, and
checking for convexity in all variables in complete generality requires
examining the positivity properties of the eigenvalues of the 
$(r+3)\times (r+3)$ matrix $\partial_A \partial_B U$. When $PV$ is zero, or negative, this matrix always has a negative eigenvalue (see discussion around equation (\ref{eq:PVStability})) and the system cannot be stable. $PV>0$ is a necessary but not sufficient condition for local thermodynamic stability.  

This question of local thermodynamic stability when $\Lambda$ is allowed to vary was addressed, for electrically neutral black holes, in \cite{Dolan:2014lea}.
A sufficient set of conditions was given 
in terms of the adiabatic compressibility at constant angular momentum,  
\beq \kappa_{J,S}=-\frac 1 V \left.\frac{\partial V}{\partial P}\right|_{J,S},
\eeq the heat capacity at constant $P$ and $\Omega$, 
\beq C_{\Omega,P}= T\left.\frac{\partial S}{\partial T}\right|_{\Omega,P},
\eeq and the adiabatic momentum of inertia tensor, 
\beq
{\cal I}_{S,P}^{ij}=\left.\frac{\partial J^i}{\partial \Omega_j}\right|_{S,P}
\eeq
(this last is automatically symmetric since it is the inverse of 
$\left.\frac{\partial \Omega_i}{\partial J^j}\right|_{S,P}$ and $\Omega_i=\left.\frac{\partial H}{\partial J^i}\right|_{S,P}$).

For electrically neutral asymptotically AdS Myers-Perry black holes there is indeed a completely
stable region provided $P$ is large enough and the rotation co-efficients are not too large.   

The stability conditions can also be expressed in terms of constant $\Omega$
rather than constant $J$: if everything is allowed to vary the question of
complete local stability is independent of the variables used, though it may be expressed differently in different variables.  For example 
in 4-dimensions, when $\Lambda=0$,  $C_J$ is negative in some regions of parameter space  and positive in others while the isothermal moment of inertia tensor,
${\cal I}_T$, has the opposite sign to $C_J$, but the product is always negative so there is no stable region.  On the other hand $C_\Omega$ is always negative
and ${\cal I}_S$ is always positive, so in $(S,\Omega)$ variables the instability is never visible in the moment of inertia.
Local stability is of course a question of constraints and stability can be
restored by fixing some variables.  For example fixing $P$
and $J$ gives the CKK critical point in 4-dimensions, with a second order
phase transition, while there is no such critical point at fixed $\Omega$. 

The above discussion is in terms of local thermodynamic stability, determined
by curvature properties of the thermodynamic potentials associated with specific solutions of Einsteins equations, but there is also the
question of global stability: for a given set of the variables $(S,P,J,Q)$ is
there another solution of Einstein's equations with a lower free energy 
$\Xi(T,P,\Omega,\Phi)$?
The Hawking-Page phase transition is a consequence of the existence of such 
a solution.  Fixing either $J$ or $Q$ eliminates the possibility of
a Hawking-Page phase transition, because AdS without a black hole cannot have non-zero angular momentum or electric charge, whereas fixing $\Omega$ does
not protect against Hawking-Page decay of black holes as AdS without a black
hole can still rotate.

There is a number of possibilities for modified gravity and/or modified electromagnetism in $D>4$.  

\begin{itemize}

\item Non-linear electrodynamics in general $D$, using the
Einstein-Born-Infeld action, was treated in \cite{Gunasekaran:2012dq,Zou:2013owa,Banerjee:2012zm}.  For charged black holes there is a critical point in any $D\ge 4$, with mean field exponents \cite{Gunasekaran:2012dq}.
The analytic form of the singularities in the heat capacity 
on the spinodal curve was determined in \cite{Banerjee:2012zm}.

Another version of non-linear electrodynamics, known as 
the power Maxwell invariant form, which replaces $-F^2$ in the Maxwell action with  
$(-F^2)^n$ for some positive integer $n$, was considered 
in \cite{Hendi:2012um,Arciniega:2014iya},
critical points with mean field exponents again emerged.

\item The effect of varying the cosmological constant on the thermodynamics of Horava-Lifshitz black holes was investigated in \cite{Sadeghi:2012ce}.

\item The $D$-dimensional Einstein-Maxwell action can be modified by adding a Gauss-Bonnet term.  The Lagrangian becomes
\beq
{\cal L}_{G-B}=\frac1{16\pi G_{(D)}}\left\{R-2\Lambda+ a^2 \Bigl(R_{\mu\nu\rho\sigma}R^{\mu\nu\rho\sigma}-4R_{\mu\nu}R^{\mu\nu}+R^2\Bigr)\right\}
-\frac 1 4   F_{\mu\nu}F^{\mu\nu},
\eeq
where $G_{(D)}$ is Newton's constant in $D$-dimensions and 
$a$ is a constant with dimensions of length.
For a charged black hole with a spherical event horizon 
there are critical points, associated with large black/small black hole
phase transitions, with mean field exponents \cite{Xu:2013zea}-\cite{Wei:2014hba}.

For $D=5$ this can happen even when the black hole is electrically neutral 
but for $D\ge6$ there must be a non-zero charge \cite{Xu:2013zea}.  In \cite{Zou:2014mha} it was found that,
when the potential $\Phi$ is held fixed rather than the electric charge,
the critical point persists only for $D=5$.
The Ehrenfest equations for phase transitions associated with
this action were checked in \cite{Mo:2014mba}.

\item Lovelock gravity \cite{Lovelock:1971}
in $D$ dimensions is an extension of the Einstein
action to include higher order terms in the Riemann tensor. It is the unique
co-variant extension that gives equations of motion that involve at most second order derivatives of the metric.
The Lagrangian for Lovelock gravity theory, including a cosmological 
constant, is given by
\beq
{\cal L}=\frac{1}{16\pi G_{(D)}}\left(R -2\Lambda +
\sum_{n=2}^{r} a_n{\cal L}_n\right)
\eeq
where $a_n$ are coupling constants, with dimensions of $(length)^{2(n-1)}$,
as before $r$ is the rank of $SO(D-1)$ and ${\cal L}_n$ are the $2n$-dimensional Euler densities,
\begin{equation}
{\cal L}_n=\frac{1}{2^n}\,\delta_{[\rho_1\sigma_1\ldots \rho_n\sigma_n]}^{[\mu_1\nu_1\ldots \mu_n\nu_n]}R_{\mu_1\nu_1}^{\hskip 15pt \rho_1\sigma_1}\ldots R_{\mu_n\nu_n}^{\hskip 15pt \rho_n\sigma_n},
\end{equation}
where $\delta_{[\rho_1\sigma_1\ldots \rho_n\sigma_n]}^{[\mu_1\nu_1\ldots \mu_n\nu_n]}$ 
represents products of $\delta$-functions totally antisymmetric in both sets of indices, (if only $a_2$ is non zero, Lovelock gravity reduces to Gauss-Bonnet gravity).   The phase space is even further extended in these theories because in principle the $a_n$ can be considered to be varying thermodynamical variables as well as $\Lambda$.
Charged black hole solutions for Lovelock gravity are known \cite{Charmousis:2008kc}
and these were investigated, in the context of varying $\Lambda$ and extended phase space, for 3rd order Lovelock gravity 
in \cite{Mo:2014qsa}-\cite{Frassino:2014pha}.
There are critical points in all dimensions $D\ge 7$, with some cases exhibiting multiple re-entrant phase transitions
between small/large/small/large black holes as well as triple points.
In \cite{Mo:2014qsa} a critical point was found $D=7$, even for electrically neutral black holes.

For 3rd-order Lovelock theory in $D\ge 7$, with $a_2^2=3 a_3$, 
neutral black holes with event horizons that have hyperbolic geometry 
have non-trivial critical exponents, \cite{Frassino:2014pha} 
$\alpha=0$, $\beta=1$, $\gamma=2$ and $\delta=3$.  This is 
an example of non-mean field critical exponents, and the same exponents are also found in
higher order Lovelock gravity \cite{Dolan:2014dkkm}.

3rd order Lovelock and Born-Infeld have also been considered in parallel \cite{Mo:2014qsa,Belhaj:2014tga}.

\item A variation of Lovelock gravity, called quasi-topological gravity, was studied in \cite{Sheykhi:2013sta} and the logic turned around to use the first law, with a $PdV$ term, to extract an expression for the entropy.

\item Other possibilities exist, such as including different kinds of matter in the action, \cite{Spallucci:2013llm}.

\end{itemize}

The positive $\Lambda$ case has in $D>4$ has so far received very little attention, though charged black holes were shown to have phase transitions in \cite{Zhang:2014zmzz}.

In conclusion there is a wide range of higher order theories of gravity
and/or non-linear electrodynamics in which first and second order phase
transitions occur, in some cases an electric charge or rotation is not
even necessary.
Equations of state in the van der Waals class seem very robust, the only
example of critical exponents that are not mean field known to date is associated with $r$-th order Lovelock gravity, with odd $r$ and a specific relation between the Lovelock co-efficients so that there only one extra independent 
parameter \cite{Dolan:2014dkkm}.  

\section{Gauge/gravity duality \label{sec:AdSCFT}}

When the cosmological constant is considered to be an independent thermodynamic
variable it is very natural to ask what its thermodynamic interpretation might be
in the boundary field theory of the AdS/CFT correspondence \cite{Maldacena:1998} (for a review see \cite{Aharony:1999ti}).
In the best understood case a 10-dimensional supergravity solution, corresponding to $AdS_5\times S^5$, relates to ${\cal N}=4$, $SU(N)$ Yang-Mills theory on
the 4-dimensional boundary of $AdS_5$, at large $N$.  In this case the 5-dimensional $\Lambda$ of $AdS_5$ is related to the radius, $L$, of $S^5$ via
\beq
\Lambda= -\frac 6 {L^2}
\eeq  
and
\beq
L^4 \sim N \ell_P^4
\eeq
with $\ell_P$ the 1--dimensional Planck length. Putting a black hole in $AdS_5$
then allows finite temperature field theory to be studied on the boundary
and the Hawking-Page phase transition in the bulk was identified with
the de-confining phase transition in the boundary gauge theory in \cite{Witten:1998ab,Witten:1998cd}.

It was first suggested in \cite{Kastor:2009} that varying $\Lambda$ in the
AdS/CFT picture should be equivalent to varying the number of colors, $N$, in the boundary Yang-Mills theory.
This was also independently suggested in \cite{Johnson:2014yja}.
The thermodynamic variable conjugate to $\Lambda$ would then act as a chemical
potential, $\mu$, for color.  This chemical potential was calculated,
for ${\cal N}=4$ SUSY Yang-Mills at large $N$,
in  \cite{Dolan:2014cja} and shown to have the general properties that
a chemical would be expected to have, for example at large $T$ it is negative and is a decreasing function of $T$.  For flat event horizons $\mu$ remains
negative at small $T$ while for spherical event horizons it can pass through
zero and become positive at a temperature that is only some 6\% below the
de-confining phase transition of the gauge theory proposed in \cite{Witten:1998cd}.

In all of the above extensions of Einstein gravity and/or electrodynamics 
there are extra dimensionful parameters introduced into the action which
affect scaling and the Smarr relation.  In higher dimensions there can also be
more dimensionful parameters in the solution, over and above the
extra angular momenta associate with $r>1$.
For example in planar black holes in $D$-dimensions one can associate 
different tensions in different tangential directions on the event horizon
and the sum of the ADM mass and tensions was shown to vanish 
in \cite{El-Menoufi:2013pza}, a result which is expected from the AdS/CFT correspondence. 

The role of $\Lambda$ in the boundary field theory in 2-D dilaton gravity 
was explored in \cite{Grumiller:2014gmns}.

\section{Conclusions \label{conclusions}}

When the cosmological constant was first introduced it was considered to
be a fixed parameter, but over the years an increasing number of scenarios
in which it might vary have been considered: for example
it can be a parameter in a solution of higher dimensional gravity or it could be determined dynamically, as in
inflationary models. 
In this context extending the thermodynamic phase space of black holes to
include the cosmological constant as a thermodynamic variable seems a very natural
move and it is perhaps somewhat surprising that it has taken so long for this step
to be taken.  The result is a much richer thermodynamics than heretofore, 
with second order phase transitions, triple points and re-entrant phase transitions. 
With the exception of higher order Lovelock gravity, all second order
phase transitions studied so far have yielded classical mean field exponents, putting these gravitational systems in the same universality class as the van der Waals gas --- indeed a solution of Einstein's equations corresponding to an energy momentum tensor with physically reasonable properties and giving exactly the van der Walls equation of state was constructed in \cite{Rajagopal:2014ewa}.  From a statistical mechanical point of view of course mean field behavior is to be expected in four or more dimensions, but until a bridge 
can be built between some microscopic quantum theory of quantum gravity and the thermodynamic
potentials of black holes it is not clear how these exponents are related to 
any putative quantum gravity theory.  Perhaps the deviation from mean field
exponents found in some models of Lovelock gravity is a hint of quantum gravity effects being reproduced by the Lovelock action,
but it is as yet too early to say.
These ideas are still in the early stages of development and we can look
forward to many more exciting developments in the future.

\begin{appendix}

\end{appendix}

\end{document}